\documentclass{article}
\usepackage{emulateapj}
\usepackage{apjfonts}
\usepackage{lscape}
\usepackage{epsfig}
\newcommand{\ltsima}{$\; \buildrel < \over \sim \;$}
\newcommand{\simlt}{\lower.5ex\hbox{\ltsima}} 
\newcommand{\gtsima}{$\; \buildrel > \over \sim \;$}
\newcommand{\simgt}{\lower.5ex\hbox{\gtsima}} 

\newcommand{\chandra}{{\emph{Chandra} }}
\newcommand{\xmm}{{XMM-\emph{Newton} }}

\newcommand{\sax}{{\emph{Beppo}SAX} }

\newcommand{\lum}{ergs~s$^{-1}$}

\newcommand{\flux}{{ergs~cm$^{-2}$~s$^{-1}$ }}
\newcommand{\nh}{cm$^{-2}$}
\newcommand{\norm}{photons~keV$^{-1}$~cm$^{-2}$~s$^{-1}$}
\renewcommand{\arcsec}{\mbox{$^{\prime\prime}$} }
\newcommand{\kev}{\,\mbox{\scriptsize keV}}
\newcommand{\mum}{\:\mu\mbox{\scriptsize m}}

\newcommand{\sorg}{Arp~299 }

\begin{document}

\title{Arp~299: a second merging system with two active nuclei?}

\author{L. Ballo\altaffilmark{1,2}, V. Braito\altaffilmark{1,3}, R. Della Ceca\altaffilmark{1}, L. Maraschi\altaffilmark{1}, F. Tavecchio\altaffilmark{1}, M. Dadina\altaffilmark{4}}

\altaffiltext{1}{INAF - Osservatorio Astronomico di Brera, via Brera 28, 20121 Milan, Italy (luballo, braito, rdc, maraschi, fabrizio@brera.mi.astro.it)}
\altaffiltext{2}{SISSA/ISAS - International School for Advanced Studies, via Beirut 4, 34014 Trieste, Italy (ballo@sissa.it)}
\altaffiltext{3}{Dipartimento di Astronomia, Universit\`a di Padova, Vicolo dell'Osservatorio 2, 35122 Padova, Italy (braito@pd.astro.it)}
\altaffiltext{4}{IASF/CNR - Sezione di Bologna, via Gobetti 101, 40129, Bologna, Italy (dadina@bo.iasf.cnr.it)}

\begin{abstract}

Recent \sax observations of Arp~299, a powerful far-IR merging starburst 
system composed of IC~694 and NGC~3690, clearly unveiled for the first time 
in this system the presence of a strongly absorbed active galactic nucleus (AGN).
However the system was not spatially resolved by \emph{Beppo}SAX. 
Here we present the analysis of archival \chandra and (for the first time) 
\xmm observations, 
which allow us to disentangle the X-ray emission of the two galaxies.
The detection of a strong $6.4\:\,$keV line in NGC~3690 clearly demonstrates 
the existence of an AGN in this galaxy, while the presence of 
a strong $6.7\:\,$keV Fe-K$\alpha$ line in the spectrum of IC~694 
suggests that also this nucleus might harbor an AGN.
This would be the second discovery of two AGNs in a merging system after 
NGC~6240.

\end{abstract}


\section{Introduction}

\sorg is a powerful merging system located at D$\;=44\:\,$Mpc 
(Heckman~et~al.~1999). 
The far-IR luminosity, 
$L_{\,43-123 \mum}=2.86 \times 10^{11}\:\,L_{\,\odot}$, dominates 
the bolometric output.
The system consists of two galaxies in an advanced merging state, NGC~3690 to 
the west and IC~694 to the east, plus a small compact galaxy to the 
northwest (Hibbard~\&~Yun~1999).
The centers of the two merging galaxies are separated by 
$\sim 22^{\prime\prime}$ (Heckman~et~al.~1999), corresponding to a 
projected distance of $4.6\:\,$kpc.
In the IR  range, IC~694 shows a compact site of activity 
in the central region, while NGC~3690 is resolved 
into a complex of sources without a clear central nucleus 
(Wynn-Williams~et~al.~1991; Alonso-Herrero~et~al.~2000).
Optical spectroscopy presented in Coziol~et~al.~(1998)
shows that IC~694 can be classified as a pure starburst galaxy, while
NGC~3690 has line properties at the borderline between starburst
and LINER.

We observed \sorg with \sax in the context of a project aimed at
investigating the starburst-active galactic nucleus (AGN) connection 
in relatively nearby 
systems, where the detection threshold should allow us to detect
AGN activity even if not dominant. 
The observations unveiled for the first time in this system a strongly 
absorbed AGN 
($N_{\mbox{\scriptsize H}} \simeq 2.5\times10^{24}\:\,$\nh, with an 
intrinsic luminosity $L_{\,0.5-100\:\,\kev} \simeq 1.9 \times
10^{43}\:\,$\lum; Della~Ceca et~al.~2002). 
At the spatial resolution of the \sax instruments, however, the system 
was not resolved, so we were unable to establish in which galaxy the 
AGN resides.

\sorg was the target of both \chandra and \xmm observations, now 
available from the archives.
Here we present the analysis of these data (those from \xmm still unpublished) and discuss the possibility 
that both galaxies in the interactive system host an AGN. 
After NGC~6240 (Komossa~et~al.~2003) this would be the second case 
of two AGNs in a merging system.
In this Letter, we assume $H_0=75$~km~s$^{-1}$~Mpc$^{-1}$ and $q_0=0.5$.

\section{Observations and data reduction}

\chandra observed \sorg on 2001~July~7 for a total of $26.5\:\,$ksec,
corresponding to a net exposure of $24.2\:\,$ksec.
The observation was performed
in FAINT mode with the target centered on the aimpoint~S1 of the ACIS-I
detector.
Screened events produced by the standard pipeline processing (evt2 files) 
have been used, while data analysis has been performed using the 
software CIAO version~2.3\footnote{See http://cxc.harvard.edu/ciao/} 
and the package 
ds9\footnote{See http://hea-www.harvard.edu/RD/saotng/}.

In this work we use the excellent spatial resolution of \chandra to have 
some indication about the origin of the X-ray emission.
We show in Fig.~\ref{fig:chandraxmm} the X-ray contours obtained from the 
\chandra ACIS-I data in the $0.5 - 2\:\,$keV (Fig.~\ref{fig:chandraxmm}\emph{a}), 
$2 - 10\:\,$keV (Fig.~\ref{fig:chandraxmm}\emph{b}), and $6.3 - 6.9\:\,$keV 
(Fig.~\ref{fig:chandraxmm}\emph{c}) energy ranges superimposed on 
the \emph{Hubble Space Telescope} (HST) Wide Field Planetary Camera~2 (WFPC2) 
image obtained with the F814W filter 
($\lambda_{\mbox{\scriptsize eff}} \simeq 8203\:\,$\AA, 
$\Delta\lambda \simeq 1758.0\:\,$\AA).

The \chandra $0.5 - 2\:\,$keV emission is clearly extended. 
Three emission knots are visible within the diffuse emission. 
The north-east and fainter knot is clearly associated with IC~694,
while the other two are associated with NGC~3690.
Also the $2 - 10\:\,$keV emission is diffuse, although in this case it is 
strongly concentrated around the three knots. 
When observed in the $6.3 - 6.9\:\,$keV energy range 
(the range where the Fe-K$\alpha$ line[s], an important spectral signature of 
an AGN, resides), 
the strongest sources are clearly localized in two regions associated with 
the two merging galaxies; moreover, the north-west knot in NGC~3690 
disappears\footnote{While this paper was in the hands of the referee, a detailed spectral and spatial analysis of these \chandra data has be published in Zezas,~Ward~\&~Murray~(2003).}.

\sorg was observed by \xmm on 2001~May~6 for a total of about
$20\,$ksec, in full frame mode and with the thin filter applied.
We used only European Photon Imaging Camera (EPIC)~pn data since the MOS data are of
insufficient quality above $6\:\,$keV. 
The \xmm data have been cleaned and processed using the 
Science Analysis Software 
(SAS version~5.4) and analyzed using standard software packages 
(FTOOLS version~4.2, XSPEC version~11.2).
Event files produced from the standard pipeline processing have been 
filtered for high-background 
time intervals, and only events corresponding to patterns $0 - 4$ have been 
used (see the \xmm Users' Handbook
\footnote{See http://xmm.vilspa.esa.es/external/xmm\_user\_support/documentation/uhb\_2.0/index.html.}); the net exposure 
time after data cleaning is $\sim 14\:\,$ksec.
The latest calibration files released by the EPIC team have been used.
We have also generated our own response matrices at the position of the 
system using the SAS tasks \emph{arfgen} and \emph{rmfgen}.
No statistically significant source variability has been detected during 
the \xmm observation.

The \xmm image in the $2 - 10\:\,$keV energy range shows two sources of 
comparable brightness corresponding to the two interacting galaxies.
To compare the EPIC-pn spectrum with that obtained by \emph{Beppo}SAX 
we extracted the source counts from a circular region of radius 
$28.5^{\prime\prime}$, including the whole merging system.
The background spectrum was extracted from a nearby 
source-free circular region of $\sim 79^{\prime\prime}$ radius.
The net count rate ($0.5 - 10\:\,$keV energy range) of the total merging 
system is $0.4822\pm0.0063\:\,$counts~s$^{-1}$; it represents about 98.8\% 
of the total counts in the source extraction region.
To perform the spectral analysis, source counts were rebinned to have 
a number of counts greater than $20$ in each energy bin.

In order to derive the spectra of the two sources separately, we selected 
two smaller regions 
centered at their X-ray centroid positions (IC~694: $\mbox{RA} = 11:28:33.9$, 
$\mbox{Dec} = +58:33:43.9$; NGC~3690: $\mbox{RA} = 11:28:29.8$, 
$\mbox{Dec} = +58:33:43.9$). 
In the case of NGC~3690 the spectrum was 
extracted from a circle of radius $18.75$\arcsec (limited in size by the 
presence of the other nucleus).
For IC~694 we used a smaller radius of $14.25$\arcsec because of the 
proximity of the CCDs' gap.
The extraction regions are shown in Fig.~\ref{fig:chandraxmm}, panel~d.
The background spectra were extracted from two source-free circular regions 
close to the individual sources of radius $32.5$\arcsec and 
$27.5^{\prime\prime}$, respectively.
The net count rate ($0.5 - 10\:\,$keV energy range) of IC~694 (NGC~3690) is 
$0.1686\pm0.0044\:\,$counts~s$^{-1}$ ($0.2291\pm0.0049\:\,$counts~s$^{-1}$) 
and represents about 97.5\% (98.3\%) of the total counts in the source 
extraction region.
Source counts have been rebinned to have a number of counts greater than $10$ 
in each energy bin.

All the models discussed here have been filtered through the Galactic 
absorption column density along the 
line of sight ($N_{\mbox{\scriptsize H, Gal}}=9.92\times10^{19}\:\,$\nh; 
Dickey~\&~Lockman~1990); 
all the errors are at 90\% confidence level for 1 parameter of interest 
($\Delta \chi^2=2.71$). 
The reported line positions refer to the source rest frame; the metallicity 
of the thermal component(s) used was fixed to the solar value.

\section{Spectral analysis}

\subsection{The whole system: comparison with previous \sax results}

\sorg was observed by \sax about 7 months after the observations of 
XMM-\emph{Newton}; the \sax MECS and \xmm EPIC-pn $2 - 10\:\,$keV fluxes 
obtained assuming a simple power-law model are comparable within the 
uncertainties.
In order to check the consistency of our previous results, the \xmm EPIC-pn 
spectrum of the whole \sorg system (see previous section) was analyzed 
jointly with the \sax data.
We fitted the data in the $0.3 - 40\:\,$keV 
energy range with the \sax best-fit model (see Della~Ceca et~al.~2002 for 
details), 
composed of a soft thermal component, a ``leaky-absorber'' model 
(an unabsorbed power law~+~an  absorbed power law with the same photon index 
$\Gamma$), and two Gaussian emission lines (at $E \sim 6.4$ and 
$E \sim 3.4\:\,$keV, respectively).
The LECS to MECS and PDS to MECS normalization factors
were allowed to vary in the range suggested by the \sax Cookbook.
\footnote{See http://ftp.asdc.asi.it/pub/software\_docs/saxabc\_v1.2.ps.gz.}

The ratio of \xmm EPIC-pn data to the best-fit model discussed above shows 
a deficit of photons at $E < 0.8\:\,$keV, which was present but not 
statistically significant in the \sax data (because of the poor LECS statistics 
at low energies).
This requires additional absorption in front of the soft thermal component, 
with a column density of 
$N_{\mbox{\scriptsize H, soft}} \sim 1.5\times10^{21}\:\,${\nh} 
(consistent with the absorption found for several Seyfert~2 galaxies with 
circum-nuclear starbursts; see Levenson,~Weaver,~\&~Heckman~2001a,~2001b). 
With this modification the overall model proposed by Della~Ceca et~al.~(2002) 
well reproduces the \xmm+ \sax data; 
the values  found for the most relevant parameters 
($N_{\mbox{\scriptsize H, hard}} \sim 2.6\times10^{24}\:\,$\nh, 
$\Gamma \sim 1.89$, $kT \sim 0.64\:\,$keV, $E \sim 6.41\:\,$keV, and 
$\chi^2/\mbox{d.o.f.}=607.1/542$) are in 
good agreement with those previously obtained.
This global modelling of \sorg is also a good fit of the \xmm EPIC-pn data 
only, with the only exception that the presence of a Gaussian line at 
$3.4\:\,$keV is not required.
The absorbed fluxes and the intrinsic (i.e., unabsorbed) luminosity are 
consistent with our previous results, confirming our earlier conclusion about 
the presence of a deeply buried AGN in the system.

\subsection{X-ray emission from the two galaxies}

Using only the EPIC-pn data, we have studied the X-ray emission 
produced by the two merging galaxies. 
Their $0.5 - 10\:\,$keV band emission can be well described by 
a thermal component~+~a power law~+~a Gaussian emission line model, with 
soft X-ray absorption in addition to the Galactic one. 
Apart from the energy and equivalent width (EW) of the emission lines 
(see below), the best-fit parameters are very similar for the two galaxies 
($kT\sim 0.66\:\,$keV, $\Gamma\sim 1.9$, and 
$N_{\mbox{\scriptsize H, soft}} \sim 1.5\times10^{21}\:\,$\nh).
Furthermore, the two galaxies contribute to the observed $2 - 10\:\,$keV 
emission of \sorg with similar intensities.

As expected, at energies lower than $2\:\,$keV the dominant contribution 
is due to the thermal emission associated with the starburst component.
The luminosity of this thermal component is 
$L_{0.5-2\:\,\kev} = 1.37 \times 10^{41}\:\,${\lum} for NGC~3690 and 
$L_{0.5-2\:\,\kev} = 1.08 \times 10^{41}\:\,${\lum} for IC~694.

\begin{figure}
\vskip -4.0 true cm
\centerline{\epsfig{figure=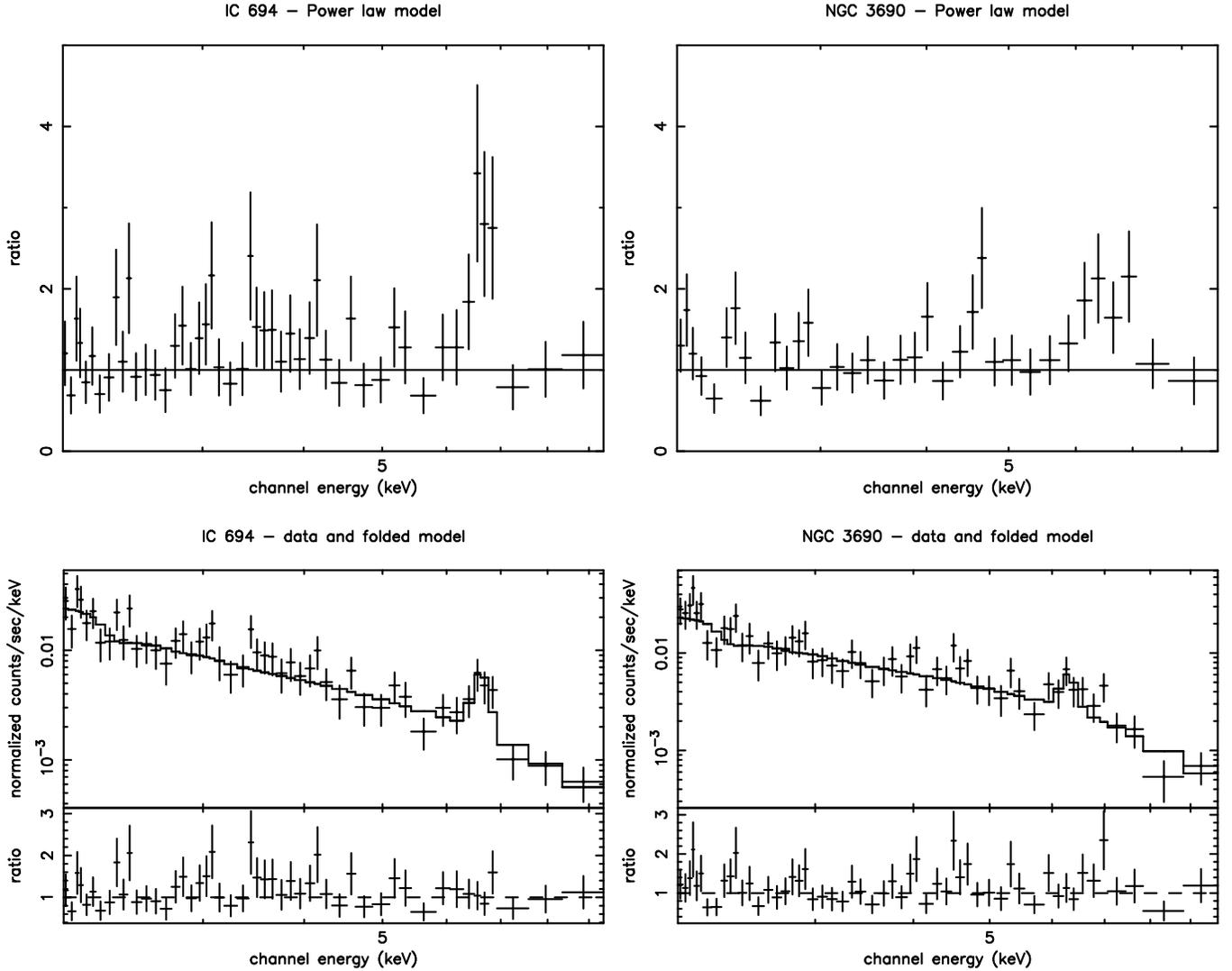, width=18cm}}
\vskip +1.0 true cm
\caption{Results of the \xmm EPIC-pn data for 
IC~694 (\emph{left}) and NGC~3690 (\emph{right}) fitted with 
different models. \emph{Top panels}: Ratio of a simple power-law 
model to the data. For demonstration purposes, data of NGC~3690 have 
been binned to have a number of counts greater than $15$ in each energy bin.
\emph{Middle panels}: Data and folded spectra for a fit with a 
power-law component and a Gaussian line.
\emph{Bottom panels}: Ratio of the data to the power law + Gaussian 
line model. 
\label{fig:plmod}}
\end{figure}

In order to study the nuclear X-ray emission of NGC~3690 and IC~694, 
we now concentrate on the $2 - 10\:\,$keV energy range, where the 
contribution from the soft thermal component is negligible.
In Fig.~\ref{fig:plmod} (\emph{top panels}) we show the ratio of the \xmm data to a single 
unabsorbed power-law fit 
($\Gamma=1.89$ for NCG~3690 and $\Gamma=1.97$ for IC~694); 
for both of these sources, the residuals suggest the 
presence of linelike features at energies between $6.3$ and 
$7\:\,$keV.
In fact, by adding a Gaussian emission line to the simple power-law model 
in both cases, the fit improves significantly according to the $F-$test; 
the results of our analysis 
are reported in Table~\ref{tab:xmmfit} and are shown in Fig.~\ref{fig:plmod} (\emph{middle and bottom panels}).
From our fit, the main difference between the two objects is the 
position and the EW of the emission lines. 
In the case of IC~694, the energy of this feature is
consistent with He-like Fe-K$\alpha$, while in the case of  NGC~3690
the energy is consistent with Fe-K$\alpha$ from neutral Fe.

\section{Discussion}
In the following section we discuss the implications of the spectral analysis 
previously described, focusing on the location of the deeply buried AGN 
in this system, as we know from \sax observation.
Because of the absorbing column density observed by \sax 
($\sim 2.5 \times 10^{24}\:\,$\nh), the direct X-ray continuum from the 
obscured AGN can be see only at energies greater than $10\:\,$keV. 
So in the energy range covered by \xmm it is completely absorbed; the only 
observable and clear  signature of this AGN is a cold~Fe-K$\alpha$ line with 
high EW, as expected if produced by transmission through the neutral 
material responsible for the absorption measured by \emph{Beppo}SAX.
Such a line is clearly detected in the \xmm spectrum of NGC~3690, strongly 
arguing for the presence in this galaxy of the absorbed AGN inferred
from the \sax observation.
Assuming that NGC~3690 produces the whole hard X-ray flux observed by 
the \sax PDS, the continuum observed by \xmm in the $2 - 10\:\,$keV 
energy range is only $\sim 2\%$ of the intrinsic one.
This continuum is probably due to a combination of emission from sources 
related to the starburst (e.g., X-ray binaries) and/or reprocessed AGN 
emission (reflection and/or scattering) to our line of sight. 
The \xmm data do not allow us to disentangle these different contributions.

\begin{figure}
\vskip -0.7 true cm
\centerline{\epsfig{figure=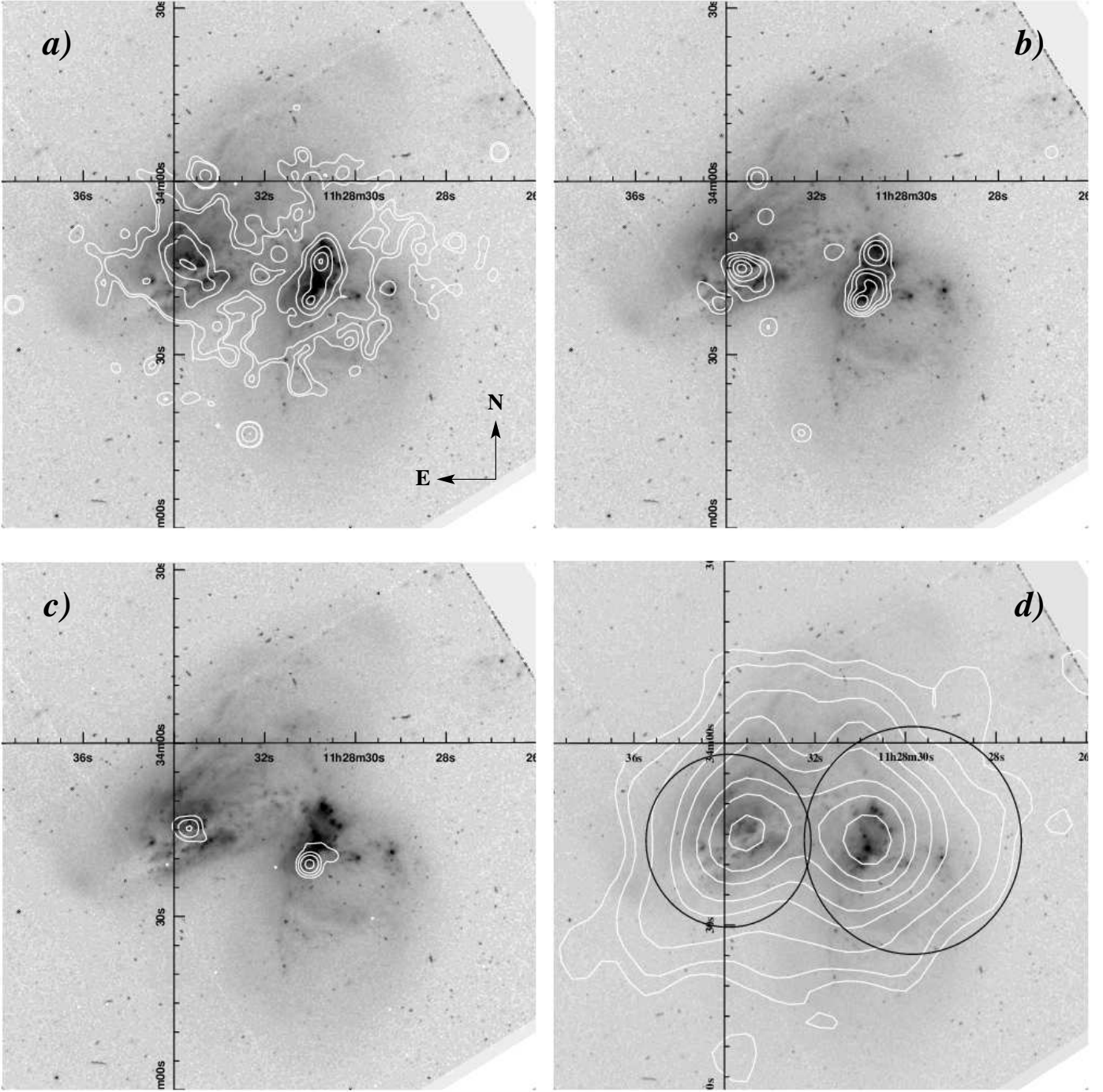}}
\vskip +0.7 true cm
\caption{X-ray contours derived (using a Gaussian smoothing point-spread function [PSF] 
with a FWHM of $1\:$pixel, i.e.,~comparable with the \chandra PSF) from the 
\chandra ACIS-I data in different energy ranges superimposed on the 
HST WFPC2 image; 
\emph{panel~a} $0.5 - 2\:\,$keV (contours corresponding to $0.13$, $0.319$, 
$0.634$, $1.264$, $3.154$, $4.729$, and $6.304\:$counts/pixel; the mean 
background is $0.004\:$counts/pixel), where we detected $284 \pm 17$, 
$468 \pm 22$, and $307 \pm 18$ net counts within a radius of $3$\arcsec 
from the emission peaks of the eastern, north-western, and south-western 
knots respectively; 
\emph{panel~b} $2 - 10\:\,$keV (contours corresponding to 
$0.24$, $0.51$, $0.96$, $1.86$, and $2.76\:$counts/pixel; the mean 
background is $0.06\:$counts/pixel), where the net 
counts close ($r < 3^{\prime\prime}$) to the emission peaks are $116 \pm 11$, 
$73 \pm 9$, and $165 \pm 13$; 
\emph{panel~c} $6.3 - 6.9\:\,$keV (contours corresponding to $0.03834$, 
$0.09534$, $0.19034$, and $0.38034\:$counts/pixel; the mean 
background is $0.00034\:$counts/pixel), 
with net counts of $6$ for the eastern source and $15$ for the western one. 
\emph{Panel~d}: \xmm EPIC-pn contours of Arp~299 in the $0.5 - 10\:\,$keV 
band, corresponding to $3\sigma$, $5\sigma$, $10\sigma$, $20\sigma$, 
$30\sigma$, $50\sigma$, $70\sigma$ and $100\sigma$, superimposed on the same 
image; the circles mark the regions considered for the spectral analysis. 
\label{fig:chandraxmm}}
\end{figure}

The case of IC~694 is more ambiguous, since the $6.7\:\,$keV line from highly
ionized Fe could be produced  by a high-temperature thermal plasma. 
In fact, we have been able to reproduce  
the $2 - 10\:\,$keV spectrum of IC~694 (continuum~+~line) 
using a combination of a cutoff power-law model 
(reproducing the integrated emission of X-ray binaries; see 
Persic~\&~Rephaeli~2002) 
and a thermal (MEKAL) model with $kT \simeq 5.5\:\,$keV; 
the two components are linked so as to reproduce the fraction of X-ray 
emission assigned by \chandra to discrete sources (Zezas~et~al.~2003).
During the fit the slope of the cutoff power-law model 
was constrained to vary between $1.3$ and $1.5$, while we fixed the cutoff 
energy to $E=8\:$keV.
The resulting temperature is higher than the values typically found in 
supernova remnants (SNRs), but consistent with that found, for instance, in the SNR N132D by 
Behar~et~al.~(2001).
Assuming a typical X-ray luminosity of young SNRs of 
$L_{\mbox{\scriptsize X}} \sim 10^{37}\:\,${\lum} and a typical duration of 
the hot phase of $1000\:\,$yr (see Persic~\&~Raphaeli~2002 for a detailed discussion on this topic), the measured 
$2 - 10\:\,$keV thermal luminosity of IC~694 
($\sim 6.5 \times 10^{40}\:\,$\lum) implies about 
$6500\:\,$SNRs in the nuclear starburst and a supernovae rate 
of $6.5\:\,$yr$^{-1}$; the latter value is about a factor of $10$ larger than 
the supernovae rate estimated in the central part of IC~694 
($< 5\arcsec$, where the bulk of the $2 - 10\:\,$keV emission is produced; 
see Fig.~\ref{fig:chandraxmm}\emph{b}) from radio and near-IR observations 
(Alonso-Herrero~et~al.~2000)\footnote{We also tried a fit leaving the 
intensity of the binary cutoff power law model free to vary. 
In this case, the supernovae rate implied by the best fit luminosity 
is $7.3\:\,$yr$^{-1}$. 
Similar results have been obtained leaving free also the slope and the 
cutoff energy.}. 
Thus, although the data do not definitively rule out the starburst origin of 
the emission line, this possibility implies rather extreme 
conditions (number of SNRs and high plasma temperature).

In spite of the fact that the luminosity of the Fe-K$\alpha$ emission 
line might depend on several ambient factors, a comparison with starburst 
galaxies showing such line can, however, supply some information about the 
expected line intensity.
We note that among ``pure" starburst galaxies, 
the He-like~Fe line at $E \sim 6.7\:\,$keV with an EW comparable to that 
of IC 694 is firmly detected only in NGC~253\footnote{The only other 
starburst galaxy that shows a line from highly ionized Fe is M82, but its 
EW is significantly lower than that measured here  
(Cappi~et~al.~1999, Rephaeli~\&~Gruber~2002).}.
In order to compare our results with those obtained with \chandra 
for NGC~253 (Weaver~et~al.~2002), we estimated the central 
($\sim 5^{\prime\prime}$) FIR luminosity of 
IC~694 using the radio measurement at $1.4\:\,$GHz (taken from the 
Faint Images of the Radio Sky at Twenty~cm [FIRST] 
survey\footnote{See http://sundog.stsci.edu/}) and 
the well-known radio/IR correlation for star-forming galaxies (Condon~1992).
We have rescaled the Fe line luminosity of NGC~253 reported by 
Weaver~et~al.~(2002) using the ratio of the FIR luminosities of the two 
galaxies, and we have found that the starburst emission could 
account for about  $20\%$ of the observed line intensity in IC~694.
Note that also NGC~253 may harbor a hidden AGN, as suggested by 
some authors (see, e.g., Mohan,~Anantharamaiah,~\&~Goss~2002).

So (as also suggested on the basis of its radio properties; see 
Gehrz, Sramek, \&~Weedman 1983), there is a strong possibility that in 
the nucleus of IC~694, an AGN may be present.
In this case, the presence of an He-like~Fe-K$\alpha$ emission 
line suggests that the AGN continuum could be scattered/reflected 
by a highly ionized gas.
A similar emission line (not accompanied by a cold Fe-K$\alpha$ line) 
has been recently found by \xmm in the 
FRI galaxy NGC~4261 (Sambruna~et~al.~2003).

Indeed we tried to model the spectrum with a reflected component 
as described by Ross \&~Fabian (1993; available in XSPEC as a 
\emph{table} model file\footnote{See 
http://heasarc.gsfc.nasa.gov/docs/xanadu/xspec/models/iondisc.html}).
Since the predicted continuum due to reflection by an ionized slab can be 
characterized by features at low energies, we considered the full 
$0.5 - 10\:\,$keV spectrum, adding a MEKAL thermal component
at low energies to take into account the starburst contribution.
This model accounts quite well for the entire spectrum 
($\chi^2/$d.o.f.$=201.6/183$), and can reproduce the Fe-K$\alpha$ line with 
physically acceptable values for the main parameters: 
$kT \sim 0.2\:\,$keV, $\Gamma \sim 2$, and an ionization 
parameter\footnote{$\xi \equiv L_{\mbox{\scriptsize ill}}/(n\,R^2)$, 
where $L_{\mbox{\scriptsize ill}}$ is the luminosity of the continuum, 
$n$ is the numerical density (part~cm$^{-3}$) of 
the illuminated slab and $R$ is the distance between the slab and the 
illuminating source.} $\xi=2.6 \times 10^3$.

The absence of a neutral Fe-K$\alpha$ line and the lack of any sign of 
absorption due to a medium with high column density indicate that the 
AGN inside IC~694 is not heavily absorbed 
($N_{\mbox{\scriptsize H}} \leq 10^{22}\:\,$\nh). 
The observed $2 - 10\:\,$keV radiation is probably the direct 
emission of the central source, an AGN of low luminosity 
($L_{\mbox{\scriptsize X}} \sim 10^{41}\:\,$\lum) surrounded by a cloud of 
highly ionized gas.
The most probable reason that prevents us from identifying IC~694 as an AGN 
from optical spectroscopic observations is the strong circum-nuclear starburst 
(note that the FIR luminosity is about 3~orders of magnitude greater 
than the X-ray luminosity of the AGN, see Charmandaris,~Stacey,~\&~Gull~2002), 
which could dilute its optical light 
(see, e.g., Georgantopoulos,~Zezas,~\&~Ward~2003).

To conclude, although a starburst origin of the X-ray emission 
observed in IC~694 cannot be ruled out, the most 
plausible hypothesis to explain the X-ray data 
presented here seems to be the existence of an AGN in each merging galaxy, one highly
absorbed ($N_{\mbox{\scriptsize H}} \simeq 2.5\times10^{24}\:\,$\nh) and of 
high luminosity ($L_{\,0.5-100\:\,\kev} \simeq 1.9 \times 10^{43}\:\,$\lum), 
the other one less luminous ($L_{\,2-10\:\,\kev} \simeq 10^{41}\:\,$\lum) 
and surrounded by highly ionized gas. 
In order to establish the AGN activity in IC~694, hard ($> 10\:\,$keV)
X-ray observations with angular resolution sufficient to 
disentangle the X-ray emission from the two galaxies would be needed. 
Such observations are far to come (Constellation~X). 
Longer and/or repeated \xmm observations, providing 
information on the variability of the X-ray sources, would be at the moment 
the only means to test the AGN hypothesis for IC~694. 

\acknowledgements

This research was based on observations obtained with XMM-\emph{Newton}, funded by ESA Member 
States and the USA (NASA).
We thank A. Celotti and M. Cappi for useful comments.
This work receive partial financial support from ASI 
(I/R/037/01, I/R/062/02, and I/R/047/02) and MURST (Cofin-2001).
We thank the referee, A. Zezas, for useful comments and suggestions that 
have substantially improved the paper.

\clearpage

\begin{landscape}
 \begin{table*}
  \begin{center}
  \caption{Results of the spectral analysis (EPIC-pn $2-10\:\,$keV): partially absorbed power law~+~narrow Gaussian line.}\label{tab:xmmfit}
\scriptsize{
   \begin{tabular}{ccccccccccccc}
    \hline
    \hline
     & & \multicolumn{2}{c}{Power Law} & & \multicolumn{3}{c}{Line} & & & & & \\
    \cline{3-4} \cline{6-8}
     & & $\Gamma$ & Norm$^{(a)}$ & & $E$ & Norm$^{(b)}$ & EW & $N_{\mbox{\scriptsize H, soft}}^{(c)}$ & Flux$^{(d)}$ & Luminosity$^{(e)}$ & $\chi^2/$d.o.f.& \\
     & & & & & (keV) & & (eV) & ($10^{21}\,$\nh) & ($10^{-13}\,$\flux) & ($10^{41}\,$\lum)& 
    \vspace{0.2cm}\\
    \hline
    \multicolumn{13}{c}{}\\
    IC~694 & & 1.95$\pm0.20$ & 1.52$^{+0.43}_{-0.32}$ & & 6.69$^{+0.12}_{-0.09}$ & 3.02$^{+1.40}_{-1.46}$ & 818$^{+380}_{-396}$ & 2.35 & 4.33 & 1.02 & 40.62/45 & \\
    \multicolumn{13}{c}{}\\
    NGC~3690$^{(f)}$ & & 1.80$^{+0.44}_{-0.32}$ & 1.29$^{+1.46}_{-0.49}$ & & 6.36$^{+0.27}_{-0.14}$ & 1.92$^{+1.20}_{-1.31}$ & 422$^{+262}_{-288}$ & 5.56 & 4.37 & 1.06 & 50.45/50 & 
    \vspace{0.2cm}\\
    \hline
    \multicolumn{13}{c}{}\\
    \multicolumn{13}{l}{\footnotesize NOTE: Errors are quoted at the 90\%
confidence level for 1 parameter of interest ($\Delta \chi ^2=2.71$).} \\
    \multicolumn{13}{l}{\footnotesize The net count rate in the $2-10\:\,$keV energy range is $0.0363\pm0.0031\:\,$counts~s$^{-1}$ for IC~694 and} \\
    \multicolumn{13}{l}{\footnotesize $0.0401\pm0.0031\:\,$counts~s$^{-1}$ for NCG~3690 (about 97.6\%  and 97.3\% of the total counts, respectively).} \\
    \multicolumn{13}{l}{\footnotesize$^{(a)}$ In units of
$10^{-4}\,$\norm\,@$1\,$keV.} \\
    \multicolumn{13}{l}{\footnotesize$^{(b)}$ In units of
$10^{-6}\,$photons~keV$^{-1}$~cm$^{-2}$~s$^{-1}$ in the line.} \\
    \multicolumn{13}{l}{\footnotesize$^{(c)}$ Column density of neutral
hydrogen in addition to $N_{\mbox{\scriptsize H, Gal}}=9.92\times10^{19}\:\,$\nh.} \\
    \multicolumn{13}{l}{\footnotesize$^{(d)}$ Observed X-ray fluxes.} \\
    \multicolumn{13}{l}{\footnotesize$^{(e)}$ Observed X-ray luminosities.} \\
    \multicolumn{13}{l}{\footnotesize$^{(f)}$ The line profile appears marginally broad ($\sigma=0.32^{+0.64}_{-0.25}\:\,$keV). This can be due to the blending of several lines: } \\
    \multicolumn{13}{l}{\footnotesize if the $2 - 10\:\,$keV continuum is produced by a ``warm mirror'' that scatters the primary radiation of the central} \\
    \multicolumn{13}{l}{\footnotesize source, then we would expect emission lines from highly ionized~Fe. A second linelike feature seems to be present} \\
    \multicolumn{13}{l}{\footnotesize at higher energies ($E \sim7 \:\,$keV), but the present statistics preclude from firm conclusions.} \\
   \end{tabular}}
  \end{center}
 \end{table*}
\end{landscape}

\end{document}